\documentclass[a4paper, amsfonts, amssymb, amsmath, reprint, showkeys, nofootinbib, twoside]{revtex4-1}

\usepackage[english]{babel}
\usepackage{amssymb}
\usepackage{amsthm}
\usepackage{amsmath}
\usepackage{mathrsfs} 
\usepackage[scr=boondox]{mathalfa}
\usepackage{multirow}
\usepackage{adjustbox}
\usepackage{float}
\usepackage{xcolor}
\usepackage{subcaption}
\usepackage{dcolumn}
\usepackage{bm}
\usepackage{multirow}
\usepackage{array}
\usepackage{siunitx} 
\usepackage{booktabs}
\usepackage{calligra}
\usepackage{colortbl}
\usepackage{graphicx}

\usepackage{xcolor,colortbl}

\definecolor{Gray}{gray}{0.85}
\definecolor{Orange}{rgb}{1,0.7,0}
\definecolor{Purple}{rgb}{0.25,0,0.78}
\definecolor{White}{rgb}{1.0,1.0,1.0}

\newcolumntype{a}{>{\columncolor{Gray}}c}
\newcolumntype{b}{>{\columncolor{White}}c}

\usepackage[utf8]{inputenc}
\usepackage[colorinlistoftodos, color=green!40, prependcaption]{todonotes}

\begin{document}
\title{Consensus effects of social media synthetic influence groups on scale-free networks}

\author{Giuliano G. Porciúncula}

    \affiliation{F\'isica de Materiais, Universidade de Pernambuco, Recife, PE 50720-001, Brazil}
    \affiliation{Department of Physics, Northeastern University, Boston, MA 02115, USA}
    
    \author{Marcone I. Sena-Junior}
    \affiliation{F\'isica de Materiais, Universidade de Pernambuco, Recife, PE 50720-001, Brazil}
    
    \author{Luiz Felipe C. Pereira}
    \affiliation{Departamento de F\'isica, Universidade Federal de Pernambuco, Recife, PE 50670-901, Brazil}
    \affiliation{Dipartimento di Fisica, Sapienza Università di Roma, Roma, RM 00185, Italy}

    \author{Andr\'e L. M. Vilela}
    \email[Corresponding author. E-mail address: ]{andre.vilela@upe.br}
    \affiliation{Data Science and Analytics, SUNY Polytechnic Institute, Utica, NY 13502, USA}
    \affiliation{F\'isica de Materiais, Universidade de Pernambuco, Recife, PE 50720-001, Brazil}
    \affiliation{Departamento de F\'isica, Universidade Federal de Pernambuco, Recife, PE 50670-901, Brazil}

\date{\today} 

\begin{abstract}
Online platforms for social interactions are an essential part of modern society. With the advance of technology and the rise of algorithms and AI, content is now filtered systematically and facilitates the formation of filter bubbles. This work investigates the social consensus under limited visibility in a two-state majority-vote model on Barab\'asi-Albert scale-free networks. In the consensus evolution, each individual assimilates the opinion of the majority of their neighbors with probability $1-q$ and disagrees with chance $q$, known as the noise parameter. We define the visibility parameter $V$ as the probability of an individual considering the opinion of a neighbor at a given interaction. The parameter $V$ enables us to model the limited visibility phenomenon that produces synthetic neighborhoods in online interactions. We employ Monte Carlo simulations and finite-size scaling analysis to obtain the critical noise parameter as a function of the visibility $V$ and the growth parameter $z$. We find the critical exponents $\beta/\bar{\nu}$, $\gamma/\bar{\nu}$ and $1/\bar{\nu}$ of and validate their unitary relation for complex networks. Our analysis shows that installing and manipulating synthetic influence groups critically undermines consensus robustness.
\end{abstract}

\keywords{Sociophysics, Phase transition, Critical phenomena, Monte Carlo simulation, Finite-size scaling}

\maketitle

Internet and social media are integral aspects of the modern lifestyle. As of April $2024$, $5.44$ billion people use the internet, representing $67.1\%$ of the world's population. In the United States, $96.9\%$ of the population uses the internet, a percentage surpassed only by Northern Europe, where $97.4\%$ of the population is online \citep{datareportal2024}. Many rely on the internet as their primary source for reading news, utilizing dedicated news websites and social media platforms. News websites supply broad coverage, while social media offers a user-curated flow of information and perspectives. This combination of journalism and social media exchange transforms how we consume news, making the internet a reliable tool for information flow.

Social media promotes personalized advertisements while engaging its audience and users via sophisticated algorithms and social filters. However, a natural consequence of such market strategy is the fabrication of a synthetic majority by amplifying certain content and viewpoints. Content filtering on social media can promote the formation of filter bubbles and polarization \citep{flaxman2016,vilela2021majority,galam2023,kweron2022}, where users are more likely to receive content that aligns with their current beliefs. This environment can isolate affected users, making them believe their or others' opinions are the most prevalent, even if this is not the case. 

A dangerous impact of a mainly profit-based advertising algorithm is the manipulation of information flow, which can distort public perception. In particular, making minority opinions appear dominant and thereby influencing social and political paradigms in ways that may not accurately reflect the broader population's views.

Facebook \citep{Facebook} and Instagram \citep{Instagram}, do not fully disclose how their content filtering and distribution algorithms operate. Facebook considers several factors when determining what content to show a user, including the posts from friends and pages the user follows, the time of day, and the user engagement statistics, such as the likelihood of the user's comments, reading time, and watch time. Additionally, Facebook evaluates the post's statistics, including the number of clicks, time people spent on the post, likes, comments, shares, and the likelihood of clickbait or links to low-quality webpages. Instagram uses different algorithms for various parts of the app, such as feed, stories, explore, reels, and search. However, these algorithms operate similarly to Facebook's, following the same structure and analyzing similar factors.

Improving social algorithms is critical to the balance of a thriving contemporary society, and opinion dynamics offers a prominent set of tools to investigate social collective phenomena. One of the simplest social models is the majority-vote model \citep{de1992isotropic}, in which individuals in a contact network interact by exchanging opinions. The model includes an agreement chance $1-q$, representing individuals' likeness to adopt the majority opinion of their peers. The individual opinion in the majority-vote model may assume one of two states: $+1$ or $-1$. The model exhibits a consensus to dissensus second-order phase transition at a finite critical noise value $q_{c}$ that depends on social network structure. The majority-vote model critical exponents in square lattices follow the Ising universality class \citep{Grinstein1985}.

Several investigations of the majority-vote model expanded the social connection structure to topologies that improve the representation of real-world networks and complex systems. Long-range interactions of small-world and spatially embedded networks enable consensus-dissensus phase transitions, in which the critical point increases monotonically with the rewiring and connection probability \citep{campos2003small, sampaio2016majority}. Random connections deeply influence the second-order phase transition of the model for average connectivity $\left< k \right> > 1$ \citep{lima2008majority, pereira2005majority}. Growth and preferential attachment of scale-free networks yield an increasing critical noise with connectivity density \citep{lima2019kinetic, lima2006majority}. Individual diffusion and permanent removal abruptly decrease consensus robustness  \citep{crokidakis2012impact}. Many other studies implement different social network scenarios that yield exuberant collective phenomena \citep{oliveira2024entropy}.

 Researchers have made significant modifications to the original majority-vote model to capture particular effects on opinion dynamics. These alterations include indirect social influence, an effect that decreases critical noise for scale-free networks with the increase of a degree-weight parameter $\alpha$ \citep{Yook2021}. Agent's independent behavior reduces the consensus as the chance of an independent decision $p$ increases \citep{vieira2016phase}. Anticonformists and strong opinioned agents were also considered, eventually promoting a spin-glass-like phase and decreasing critical noise, respectfully \citep{krawiecki2018spin, vilela2018effect}. An additional number of available opinions generally promote a higher critical noise value necessary to dismantle consensus \citep{brunstein1999universal, chen2015critical, lima2012three, tome2002cumulants, vilela2020three, melo2010phase}. The majority-vote opinion dynamics has also been successfully applied to financial market modeling, versatilely capturing complex dynamics of such systems \citep{granha2022opinion, lima2010analysing}.

The fundamental properties enabling the effectiveness of social media engagement algorithms rely on the topological characteristics of the underlying network connecting users. In social media, new users initially connect with their real-world acquaintances and subsequently receive algorithmic suggestions to follow users with many connections. Extensive investigations of real-world internet networks, such as the World Wide Web, have shown that these networks exhibit a degree distribution that follows a scale-free power law. This scale-free property arises from the continuous expansion of a network in which new nodes preferentially attach to existing nodes with a high number of connections \citep{adamic2000power, albert2002statistical, barabasi1999emergence, Yook2002}. 

The Barabási-Albert algorithm to build scale-free networks includes growth and preferential attachment as critical mechanisms to reproduce the power law degree distribution and enable the emergence of influential users, known as hubs. Hub users connect with many other network users, have an extensive reach, and are heavily present on online social media. Within these considerations, scale-free networks provide an essential framework for representing social media.

In order to investigate the effects of filtering in social media on consensus formation, we study the majority-vote model with a limited visibility parameter on scale-free networks. Our Monte Carlo simulations reveal phase transitions depending on the noise parameter $q$, visibility parameter $V$ and network topology, characterized by the growth parameter $z$ and the total number of nodes. Additionally, we estimate the critical exponents using finite-size scaling and verify their unitary relation \citep{vilela2020three}.

\begin{figure}[]
    \centering
    \includegraphics[width=0.45\textwidth]{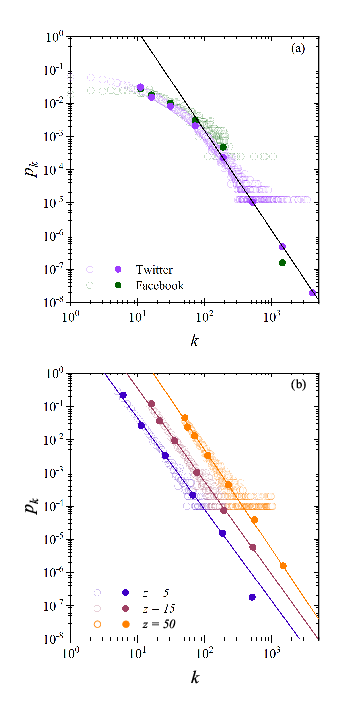}
    \caption{(a) Degree distribution of Twitter (purple) and Facebook (green) social online networks. The transparent hollow circles represent the degree distribution, while the filled circles represent the binned data. (b) Barab\'asi-Albert scale-free networks with growth parameters $z = 5$, $15$, and $50$ with $N = 10^4$. Lines follow a power law $p_{k} \sim k^{-\lambda}$ with $\lambda \simeq 3$.}
    \label{RNBAPlotSameScale}
\end{figure}

\section{Social Networks}
\label{sec:BANet}

Billions of people use social media in a highly complex network of interactions, with vast availability of data to collect and analyze. We examine Facebook and Twitter data from 2012 \citep{leskovec2012learning,TwitterData2012,FacebookData2012}. Figure \ref{RNBAPlotSameScale} (a) shows the degree distribution for the Facebook and Twitter data sets, which exhibit fat tails and scale-free properties. From $k \approx 72$ onward, the distribution follows a power-law $p_{k} \sim k^{-\lambda}$ with $\lambda = 3$, characteristic of the scale-free behavior and represented by the black line.

We evaluate critical characteristics from the data sets of the Facebook and Twitter social media networks. The Facebook network sample analyzed presents $N_F = 4039$ nodes, $L_F = 88234$ edges connecting the nodes, an average connectivity of $\left< k \right>_F \approx 43.69$, an average clustering coefficient $\left< C \right>_F \approx 0.6055$, and a diameter of $d_F = 8$. Twitter data sample presents $N_T = 81,306$ nodes, $L_T = 1342296$ edges, $\left< k \right>_F \approx 33.02$, and average clustering coefficient $\left< C \right>_T \approx 0.5653$. The Twitter network diameter is $d_F = 7$. Our results are compatible with previous investigations of such network data \cite{TwitterData2012,FacebookData2012}.

Inspired by the power law properties of real-world online networks, we model social connections using Barabási-Albert scale-free networks. Each node represents an individual user and each edge represents a social connection. We utilize two parameters to specify network design: the growth parameter $z$, indicating the number of connections each newly added node establishes within the network, and the total number of individuals $N$. We start with a complete graph of $z+1$ nodes and insert new nodes into the network until it reaches $N$ nodes in total. At each growth step, we evaluate the overall connectivity of the network and connect a new node $i$ to node $j$ with a probability that depends on $k_{i}$, the connectivity of node $i$, that is, $p_{i} = k_{i}/\sum_{\ell} k_{\ell}$.

\begin{figure*}[htb]
    \centering
\includegraphics[width=0.75\textwidth]{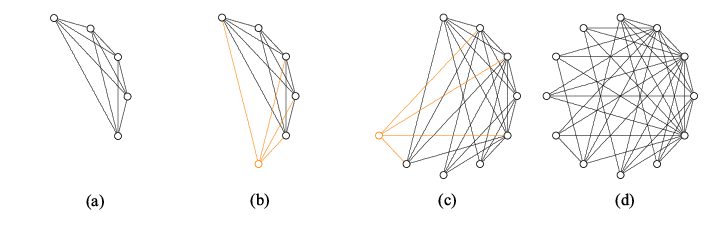}
    \caption{Barabási-Albert scale-free network formation snapshots with growth parameter $z = 4$ and volume $N = 12$, with new nodes and links added in orange.}
    \label{BAPlot}
\end{figure*}

Figure \ref{BAPlot}(a)-(d) illustrates the Barab\'asi-Albert scale-free building process for $z = 4$ and system volume $N = 12$, while Figure \ref{RNBAPlotSameScale} (e)  presents the degree distribution plot for $z = 5, 15$, and $50$. We observe a small fraction of nodes with a high number of links identified as hubs. The lines follow the degree distribution using power law with exponent $\lambda \simeq 3$, characteristic of the scale-free property \citep{adamic2000power}.

\begin{equation}
\left< k \right> = 2z-\frac{z(z+1)}{N}.
\label{kmed}
\end{equation}
For instance, based on this expression, we find $\left< k \right> (z = 15) = 29.976$, compatible with the average connectivity obtained from the Facebook and Twitter network data considered.

\section{Model Description}
\label{sec:MVMV}
An individual $i$ holds one of two opinions at any given time, $\sigma_{i} = +1$ or $\sigma_{i} = -1$, and by interacting with their neighbors, an individual has a $1 - q$ probability of agreeing with the majority opinion and a probability of disagreeing $q$. The $q$ parameter, known as social noise or social temperature, represents the ease with which an agent adopts the opposite opinion of his peers.

We model the social media filtering algorithms through a visibility parameter $V$ that limits the user's perception of its peers \citep{vilela2021majority}. During opinion formation, individuals examine their visible connections to evaluate the majority opinion with probability $0 < V \leq 1$. That is, for $V = 1$, individuals consider the opinion of all of their connected neighbors; conversely, for $V = 0$, they ignore the opinion of all their neighbors. In the particular case where an individual sees no neighbors, their opinion remains unchanged. Hence, we define the following visibility index $I(V)$ as

\begin{equation}
I(V) =
\begin{cases}
 1, & \text{with probability } V; \\ 
 0, & \text{with probability } 1-V,
\end{cases}
\label{vparameter}
\end{equation}
to enable management over the chance of evaluating a user's visible neighborhood. For a visible neighborhood, the opinion of an individual $\sigma_i$ flips with probability

\begin{multline}
    w(\sigma_i) = \frac{1}{2}\textrm{stp}\left (\sum_{\delta = 1}^{k_i}I_{i+\delta}(V)\right) \times \\ \times \left\{1-(1-2q) \, \sigma_i \, \textrm{sgn}\left[\sum_{\delta = 1}^{k_i}I_{i+\delta}(V) \, \sigma_{i+\delta}\right]\right\}
    \label{probw}
\end{multline}
\noindent where the sum runs over all $k_i$ neighbors connected to the individual $i$ and $\textrm{sgn}(x) = +1, 0, -1$ for $x < 0$, $x = 0$, and $x > 0$, respectively. The function $\textrm{stp}(y) = 1$ for $y>0$ and $0$ for $y = 0$. The visibility index of each neighbor connected to $\sigma_i$, $I_{i+\delta}(V)$, acts dynamically, being tested each time a connected neighbor is visited. If $I_{i+\delta}(V) = 0$ for all $k_i$ at a given time, then the opinion $\sigma_i$ remains unchanged as $w(\sigma_i) = 0$. Our model differs from the usual site or bond dilution since no agents are removed from the network \cite{vilela2021majority}.

\section{Visibility effects on consensus phenomena}
Generally, the majority-vote dynamics for social systems exhibit different prevailing opinion phases that depend on noise variable $q$ and other topological factors. To analyze the system's social order, we define an instantaneous order parameter as
\begin{equation}
    \mathscr{o} = \frac{1}{N}\left |\sum_{i=1}^{N}\sigma_{i} \right |,
\end{equation}
where $\sigma_i$ is the opinion variable of individual $i$ and $N$ is the network volume. To explore the system's profile over the stationary regime, we evaluate the opinionization as the average of the order parameter
\begin{equation}
    \mathscr{O}_N(q,V,z) = \left< \left< \mathscr{o} \right>_t \right>_c.
\end{equation}
We also measure the variance of the opinionization referred to as noise sensitivity
\begin{equation}
    \chi_N(q,V,z) = N\left[ \left< \left< \mathscr{o}^{2} \right>_t \right>_c - \left< \left< \mathscr{o} \right>_t \right>_c^2 \right].
\end{equation}
Analogous to the magnetic susceptibility, the noise sensitivity allows us to identify noise values that yield expressive impact on social opinion order. Additionally, to precisely estimate critical noise values that change the system's social order under different scenarios of $V$, $z$ and $N$, we evaluate the Binder fourth-order cumulant \citep{binder1981finite}, defined as
\begin{equation}
    U_N(q,V,z) = 1 - \frac{\left< \left< \mathscr{o}^{4} \right>_t \right>_c}{3 \left< \left< \mathscr{o}^{2} \right>_t \right>_c^2.}.
\end{equation}
All the averages are time averages in the stationary regime and configurational averages over independent realizations. The opinionization, noise sensitivity, and Binder cumulant depend on social noise $q$, visibility parameter $V$, growth parameter $z$, and network volume $N$. The majority-vote model with limited visibility on scale-free networks exhibits phase transitions as $q$, $z$, and $V$ change, with finite-size scaling criticality \citep{vilela2020three,stanley1971phase}.

\section{Results}
\label{sec:Results}

\begin{figure*}[htb]
    \centering
    \includegraphics[width=1\textwidth]{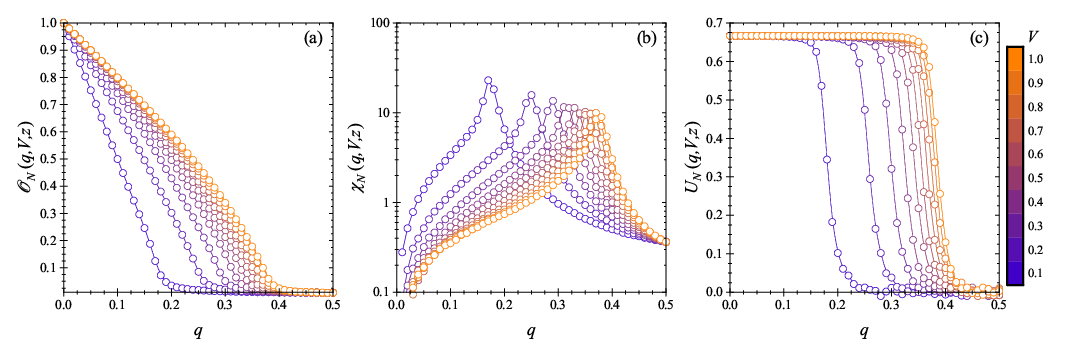}
    \caption{Visibility effects on social order for the (a) Opinionization, (b) noise sensitivity, and (c) Binder cumulant versus noise parameter $q$ for several visibility values $V$ for $z = 10$ and network volume $N = 10^4$. The lines are guides to the eye.}
    \label{MXU_Overview_z10_N10000_V0.10to1.00}
\end{figure*}

\begin{figure*}[htb]
    \centering
    \includegraphics[width=1\textwidth]{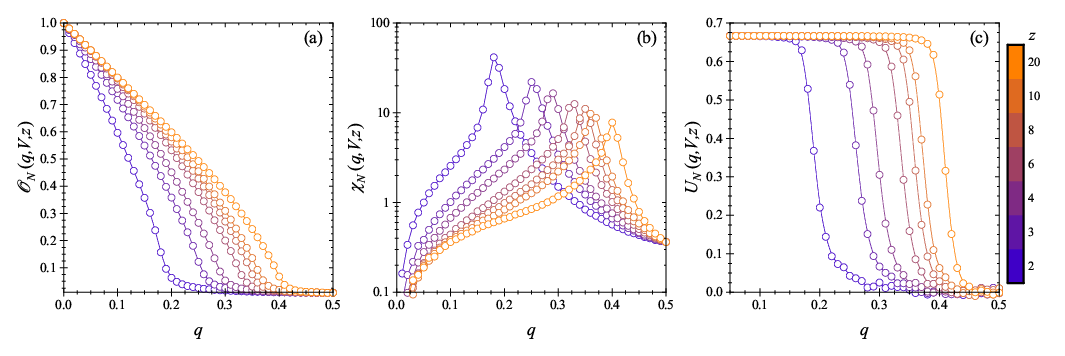}
    \caption{Growth parameter and social interaction structure effects on (a) opinionization, (b) noise sensitivity, and (c) Binder cumulant versus $q$ for several values of the growth parameter $z$ for $V = 0.80$ and $N = 10^4$ individuals. The lines are guides to the eye.}
    \label{MXU_Overview_V0.80_N10000_zvar}
\end{figure*}

In Fig. \ref{MXU_Overview_z10_N10000_V0.10to1.00}, we display the effects of different values of visibility parameter on (a) opinionization, (b) noise sensitivity, and (c) Binder cumulant as a function of the social noise $q$. We show the results for scale-free networks with growth parameter $z = 10$ and system volume $N = 10^4$. We increase the visibility parameter from $0.1$ to $1.0$ with $0.1$ increments. The social system transitions from consensus to dissensus as the noise parameter $q$ increases. We observe a macroscopic state where $\mathscr{O} \sim 1$ for small noise values, where most individuals adopt the same opinion. As $q$ increases, $\mathscr{O} \sim 0$, different opinion groups emerge, yielding a disordered phase.

We observe that the lower the visibility parameter, the lower the noise value necessary to cause a disordered phase to arise. This result shows that a synthetic influential neighborhood induces a stronger or weaker consensus phase in the social system. In Fig. \ref{MXU_Overview_z10_N10000_V0.10to1.00}(b), we plot the noise sensibility that exhibits a maximum near the critical value of $q$ that depletes consensus, that decreases with $V$. Binder cumulant from Fig. \ref{MXU_Overview_z10_N10000_V0.10to1.00}(c) also indicates a sharp decline near the noise parameter value $q$ that transitions the system from the ordered to a disordered phase. Our results indicate that social media content filtering, represented by lower visibility parameters, weakens the consensus among individuals.

Figure \ref{MXU_Overview_V0.80_N10000_zvar} shows the influence of network social structure on the social order versus $q$ with $10^4$ individuals and visibility parameter at $80\%$. In this result, we plot different values of the growth parameter, from left to right: $z = 2$, $3$, $4$, $6$, $8$, $10$ and $20$. Increasing growth parameter values alter consensus robustness regarding noise levels, in which higher values of $q$ are necessary to weaken consensus to a state where the opinionization $\mathscr{O}$ tends to zero. Figures \ref{MXU_Overview_V0.80_N10000_zvar}(b) and \ref{MXU_Overview_V0.80_N10000_zvar}(c) correspondingly demonstrate that a denser network, with a higher growth parameter $z$, promotes individual exposure to more opinions via social links, strengthening consensus.

\begin{figure*}[htb]
    \centering
    \includegraphics[width=1\textwidth]{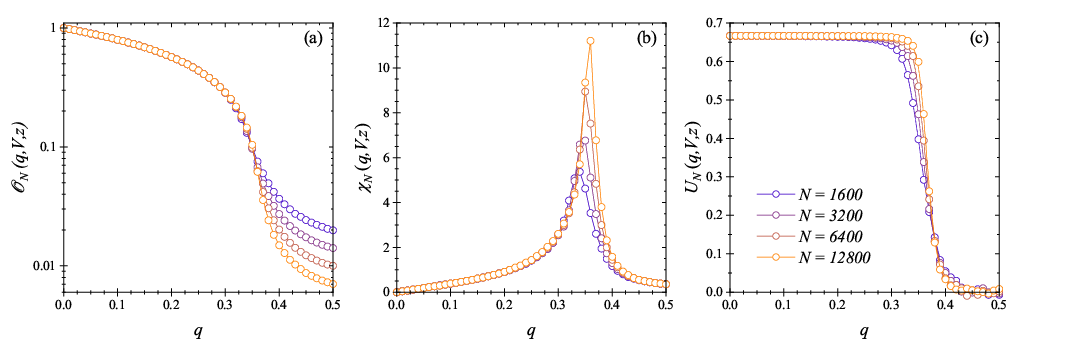}
    \caption{Finite size effects on social phase transitions for the (a) opinionization, (b) noise sensitivity, and (c) Binder cumulant versus $q$ for several values of the network size $N$ and fixed growth parameter $z = 10$ and visibility parameter $V = 0.70$. The lines are guides to the eye.}
    \label{MXU_Overview_V0.70_z10_Nvar}
\end{figure*}

System size critically affects the social consensus with dynamical visibility filtering. In Figure \ref{MXU_Overview_V0.70_z10_Nvar}, we show the (a) opinionization, (b) noise sensibility and (c) Binder cumulant as a function of social noise $q$ and $N = 1600, 3200, 6400$ and $12800$. Our results explore scale-free networks with growth parameter $z = 10$ and visibility parameter $V = 70\%$. Near the phase transition, we observe that the expressive changes in opinionization and noise sensitivity become sharper with increasing $N$, denoting that higher intensity transition phenomena emerge the bigger the connected society. Binder cumulant denotes an intersection point for different system volumes, allowing us to estimate the critical noise value $q_c$ in the thermodynamic limit $N \to \infty$.

\begin{figure}[htb!]
\centering
\includegraphics[width=0.45\textwidth]{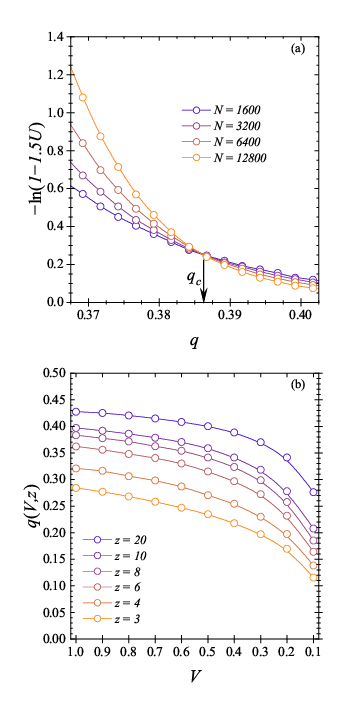}
\caption{(a) Binder cumulant as a function of $q$ near the critical noise $q_{c}$ for values of several values of network volume $N$, growth parameter $z = 10$ and visibility parameter $V = 0.8$. All curves intersect at the critical point $q_{c} = 0.3863(1)$. (b) Phase diagram $q - V$ of the majority-vote model with limited visibility on Barab\'asi-Albert networks. The solid line is just a guide to the eye. In all plots, the error bars are smaller than the symbol size.}
\label{BinderPhase}
\end{figure}

The Binder cumulant allows us to estimate critical noise values that are dependent on the growth and visibility parameters, $z$ and $V$, respectively. Figure \ref{BinderPhase}(a) illustrates the crossing point that denotes the critical temperature for $z = 10$, $V = 80\%$ and four system volumes $N = 1600, 3200, 6400$ and $12800$. Figure \ref{BinderPhase}(b) shows the phase diagram with boundaries that separate consensus-dissensus phases in opinion dynamics in the presence of synthetic majorities. Above and below the lines, the society presents ordered and disordered opinion phases, respectively. We observe that $q_{c}(V, z)$ decreases as $V$ decreases, comparable to the behavior observed in the square lattice \citep{vilela2021majority}, and increases as $z$ increases, similar to the behavior observed in the Barabasi-Albert Network \citep{vilela2020three}. The phase diagram allows us to remark that lessening synthetic neighborhoods induces societal disagreement while increasing connectivity bolsters societal consensus.

Near the phase transition point opinionization $\mathscr{O}_N(q, V, z)$, noise sensitivity $\chi_N(q, V, z)$ and Binder cumulant $U_N(q, V, z)$ satisfy the following finite-size volumetric scaling relations \citep{vilela2020three}
\begin{equation}
    \mathscr{O}_N(q,V,z)
    =
    N^{-\beta/\Bar{\nu}}
    \widetilde{\mathscr{O}}
    \left[(q-q_{c}) N^{1/\bar{\nu}}\right],
    \label{mageq}
\end{equation}
\begin{equation}
    \chi_N(q,V,z)
    =
    N^{\gamma/\bar{\nu}}
    \widetilde{\chi}
    \left[(q-q_{c}) N^{1/\bar{\nu}}\right],
    \label{suseq}
\end{equation}
\begin{equation}
    U_N(q,V,z)
    =
    \widetilde{U}
    \left[(q-q_{c}) N^{1/\bar{\nu}}\right],
    \label{bineq}
\end{equation}
where $q_{c}$ is the critical noise and $\beta/\Bar{\nu}$, $\gamma/\Bar{\nu}$, and $1/\Bar{\nu}$ are the critical exponent ratios. $\widetilde{\mathscr{O}}$, $\widetilde{\chi}$ e $\widetilde{U}$ are scalar functions that depend only on the scaled variable $x = (q-q_{c}) N^{1/\Bar{\nu}}$.
The exponent ratios $\beta/\bar{\nu}$ and $\gamma/\bar{\nu}$ are related by the unitary relation, which is derived from the hyperscaling relation and defined as \citep{vilela2020three}
\begin{equation}
    \frac{2\beta}{\bar{\nu}} + \frac{\gamma}{\bar{\nu}} = 1,
    \label{UnitRel}
\end{equation}
which we use to validate our critical analysis. We can estimate the critical exponents from the finite-size scaling relations (\ref{mageq}), (\ref{suseq}), and (\ref{bineq}). We write for $q = q_{c}$
\begin{equation}
    \ln\left[\mathscr{O}_N(q_{c},V,z)\right]
    \sim -\frac{\beta}{\bar{\nu}}
    \ln\left[N\right],
\end{equation}
\begin{equation}
    \ln\left[\chi_N(q_{c},V,z)\right]
    \sim
    \frac{\gamma}{\bar{\nu}}
    \ln\left[N\right],
\end{equation}
\begin{equation}
    \ln\left[U'_N(q_{c},V,z)\right]
    \sim
    \frac{1}{\bar{\nu}}\ln\left[N\right],
\end{equation}
where
\begin{equation}
    U'_N(q_{c},V,z)
    = -\left.\frac{\partial}{\partial q}U_N(q,V,z)\right|_{q=q_{c}}
\end{equation}    

\begin{figure*}[htb]
    \centering
    \includegraphics[width=1\textwidth]{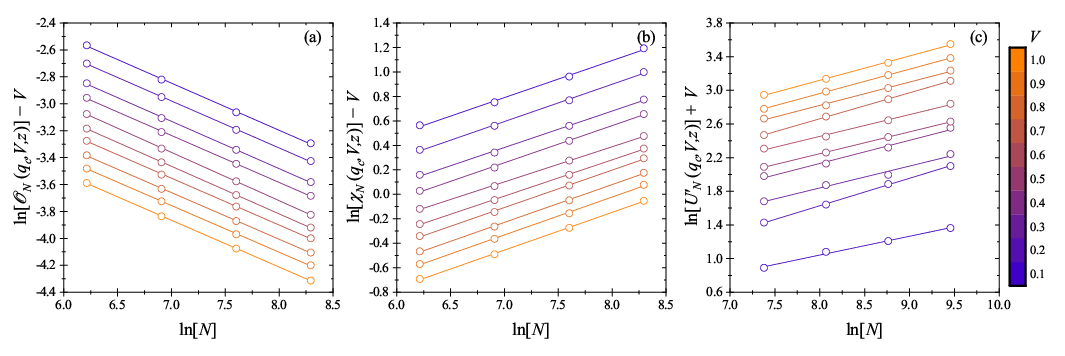}
            \caption{Natural log of (a) the Order Parameter, (b) the noise sensitivity, and (c) the derivative of the Binder versus $\ln[N]$ at the critical noise $q = q_{c}$ for several values of $V$ and fixed growth parameter $z = 10$. The lines are obtained from linear regression of the data. The critical exponents (a) $\beta/\Bar{\nu} \approx 0.35$, (b) $\gamma/\Bar{\nu} \approx 0.30$, and (c) $1/\Bar{\nu} \approx 0.28$ are obtained from the slopes of the curves.}
    \label{exponents}
\end{figure*}

The expressions are linear in $\ln[N]$, allowing us to estimate the critical exponents from the slopes of the lines. Figure \ref{exponents} presents the logarithm of the Order Parameter, noise sensitivity and Binder cumulant at $q = q_{c}$ versus the logarithm of the network size $N$. In this figure, the lines are results for $z = 10$ with $V$ from $0.1$ to $1.0$ with increments of $0.1$. We shift each line vertically to avoid overlapping. From the slope of the linear regression of the data, we obtain the critical exponents $\beta/\Bar{\nu} \approx 0.35$, $\gamma/\Bar{\nu} \approx 0.30$, and $1/\Bar{\nu} \approx 0.28$, slightly independent of the visibility parameter $V$.

\begin{figure}[htb]
    \centering
    \includegraphics[width=0.45\textwidth]{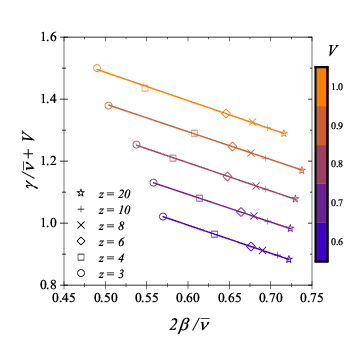}
            \caption{Characteristic unitary line $2\beta/\bar{\nu} + \gamma/\bar{\nu} = 1$ plot for several values of the growth parameter $z$ and visibility $V$. The error bars are smaller than the symbol size.}
            \label{UnRel}
\end{figure}

Table \ref{exptable} lists the critical exponents $\beta/\Bar{\nu}$, $\gamma/\Bar{\nu}$, $1/\Bar{\nu}$, and the unitary relation $2\beta/\Bar{\nu} + \gamma/\Bar{\nu}$ for $z = 4$, $10$, $20$ and values of the visibility parameter $V$ from $0.1$ to $1.0$. We highlight that for lower values of the growth parameter, critical exponents vary expressively while tending to remain indifferent to the visibility parameter as $z$ increases. This result indicates that the critical effects of limited visibility may be attenuated for social networks with high connectivity. Figure \ref{UnRel} shows the characteristic unitary line of the model for several values of growth parameter $z$ and visibility parameter $V$. The values are shifted vertically proportional to $V$. The plot shows the values of $\gamma/\Bar{\nu}$ versus $2\beta/\Bar{\nu}$. The linear fit of each curve yields a linear coefficient of approximately $1$, as expected from Eq. \eqref{UnitRel}. Our results substantiate the unitary relation holding for the majority-vote model with limited visibility on Barabási-Albert networks \citep{Yook2021, vilela2020three}.

\begin{figure*}[htb]
    \centering
    \includegraphics[width=1\textwidth]{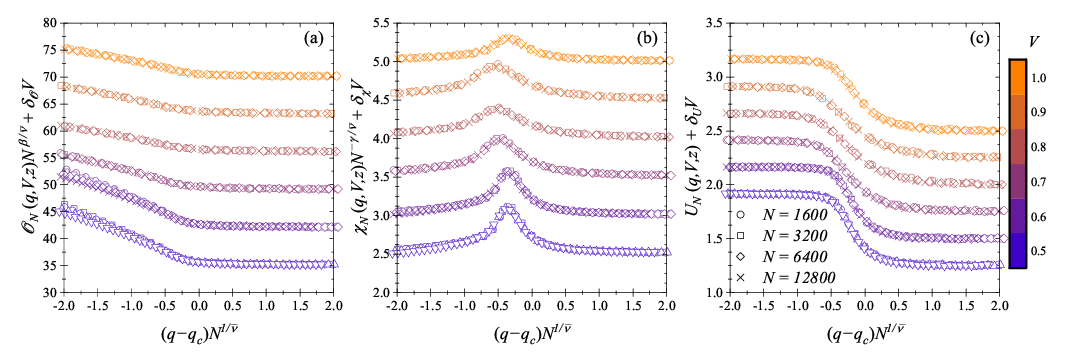}
            \caption{Data collapse using the functions (a) $\mathscr{O}_N(q,V,z)N^{\beta/\bar{\nu}}$, (b) $\chi_N(q,V,z)N^{-\gamma/\bar{\nu}}$ and (c) $U_N(q,V,z)$ versus $(q-q_{c}) N^{1/\bar{\nu}}$ for $z = 4$ and $V$ varying from $0.1$ to $1.0$ with increments of $0.1$. Each curve comprises the network volumes $N = 1600$, $3200$, $6400$, and $12800$. To avoid overlaps for different values of $V$, we shift the date sets using $\delta_{\mathscr{O}} = 75$, $\delta_{\chi} = 7$, and $\delta_{U} = 5$. The error bars are smaller than the symbol size.}
    \label{datacollapse}
\end{figure*}

\begin{table}[htb]
\begin{tabular}{ l  a  b  a  b  a  b }
\hline\hline
$z$                    & $V$   & $\beta/\bar{\nu}$ & $\gamma/\bar{\nu}$ & $1/\bar{\nu}$ & $2\beta/\bar{\nu} + \gamma/\bar{\nu}$ \\ \arrayrulecolor{white}\hline\hline
\multirow{10}{*}{$4$}  & $0.1$ & $0.37(1)$  & $0.26(2)$  & $0.21(5)$   & $1.00(3)$ &  \\ %
                       & $0.2$ & $0.360(8)$ & $0.28(1)$  & $0.13(4)$   & $1.00(2)$  &\\ %
                       & $0.3$ & $0.353(7)$ & $0.30(1)$  & $0.23(2)$   & $1.01(2)$  &\\ %
                       & $0.4$ & $0.356(8)$ & $0.30(1)$  & $0.27(2)$   & $1.01(2)$  &\\ %
                       & $0.5$ & $0.316(6)$ & $0.36(1)$  & $0.281(7)$  & $0.99(2)$  &\\ %
                       & $0.6$ & $0.316(6)$ & $0.364(6)$ & $0.282(9)$  & $1.00(1)$  &\\ %
                       & $0.7$ & $0.307(4)$ & $0.380(7)$ & $0.332(2)$  & $0.99(1)$  &\\ %
                       & $0.8$ & $0.291(3)$ & $0.409(6)$ & $0.36(1)$   & $0.991(8)$ &\\ %
                       & $0.9$ & $0.304(1)$ & $0.391(4)$ & $0.3608(7)$ & $0.999(4)$ &\\ %
                       & $1.0$ & $0.274(4)$ & $0.436(7)$ & $0.324(6)$  & $0.98(1)$  &\\ \arrayrulecolor{white}\hline\hline
\multirow{10}{*}{$10$} & $0.1$ & $0.350(6)$ & $0.30(1)$  & $0.22(1)$   & $1.00(2)$  &\\  
                       & $0.2$ & $0.349(4)$ & $0.305(8)$ & $0.326(6)$  & $1.00(1)$  &\\  
                       & $0.3$ & $0.352(5)$ & $0.298(8)$ & $0.26(2)$   & $1.00(1)$  &\\  
                       & $0.4$ & $0.349(4)$ & $0.303(7)$ & $0.27(2)$   & $1.00(1)$  &\\  
                       & $0.5$ & $0.359(3)$ & $0.287(5)$ & $0.259(4)$  & $1.005(8)$ &\\  
                       & $0.6$ & $0.354(2)$ & $0.296(4)$ & $0.26(1)$   & $1.004(6)$ &\\  
                       & $0.7$ & $0.348(3)$ & $0.306(6)$ & $0.309(4)$  & $1.002(8)$ &\\  
                       & $0.8$ & $0.347(3)$ & $0.309(5)$ & $0.28(1)$   & $1.003(8)$ &\\  
                       & $0.9$ & $0.347(4)$ & $0.310(7)$ & $0.289(2)$  & $1.00(1)$  &\\  
                       & $1.0$ & $0.348(2)$ & $0.308(4)$ & $0.289(8)$  & $1.004(6)$ &\\
                          \arrayrulecolor{white}\hline\hline
\multirow{10}{*}{$20$} & $0.1$ & $0.350(3)$        & $0.301(5)$         & $0.307(9)$    & $1.001(8)$                            &\\  
                       & $0.2$ & $0.357(3)$ & $0.287(5)$ & $0.288(8)$  & $1.001(8)$ &\\  
                       & $0.3$ & $0.357(3)$ & $0.288(7)$ & $0.289(6)$  & $1.002(9)$ &\\  
                       & $0.4$ & $0.366(3)$ & $0.274(5)$ & $0.27(1)$   & $1.006(8)$ &\\  
                       & $0.5$ & $0.366(2)$ & $0.274(4)$ & $0.295(6)$  & $1.006(6)$ &\\  
                       & $0.6$ & $0.361(3)$ & $0.283(6)$ & $0.28(1)$   & $1.005(8)$ &\\  
                       & $0.7$ & $0.362(3)$ & $0.282(6)$ & $0.294(8)$  & $1.006(8)$ &\\  
                       & $0.8$ & $0.365(3)$ & $0.278(6)$ & $0.273(8)$  & $1.008(8)$ &\\  
                       & $0.9$ & $0.369(2)$ & $0.270(4)$ & $0.287(6)$  & $1.008(6)$ &\\  
                       & $1.0$ & $0.358(4)$ & $0.289(6)$ & $0.312(5)$  & $1.00(1)$  &\\ \arrayrulecolor{black}\hline
\end{tabular}
\caption{Critical exponents $\beta/\Bar{\nu}$, $\gamma/\Bar{\nu}$, $1/\Bar{\nu}$, and the unitary relation $2\beta/\Bar{\nu} + \gamma/\Bar{\nu}$ for different values of growth parameter $z$ and visibility parameter $V$.}
\label{exptable}
\end{table}
From the finite-size scaling relations (\ref{mageq}), (\ref{suseq}), and (\ref{bineq}), we rescale measurements to obtain the universal functions $\mathscr{O}_N(q,V,z)N^{\beta/\bar{\nu}} = \widetilde{\mathscr{O}}\left[x\right]$, $\chi_N(q,V,z)N^{-\gamma/\bar{\nu}} = \widetilde{\chi}\left[x\right]$ and $
    U_N(q,V,z) = \widetilde{U}\left[x\right]$ where $x = (q-q_{c})N^{1/\bar{\nu}}$. Figure \ref{datacollapse} shows the data collapse using the universal functions for $z = 4$, where symbols represent different network sizes $N =1600$, $3200$, $6400$, $12800$, where curves have visibility $V$ ranging from $0.5$ to $1.0$. For each growth and visibility parameter, we employ our estimations for the critical noise $q_{c}$ and for critical exponents $\beta/\bar{\nu}$, $\gamma/\bar{\nu}$, and $1/\bar{\nu}$. In this result, we use $\delta_{\mathscr{O}} = 75$, $\delta_{\chi} = 7$, and $\delta_{U} = 5$, for clarity to shift the curves and avoid overlaps. Adequate collapse of the data validates our critical exponent estimations.

\section{Conclusion and final remarks}
\label{sec:Conc}
We investigate the effects of content filtering algorithms on creating synthetic neighborhoods in social networks. We model social interactions using the Barabási-Albert scale-free networks with growth parameter $z$ and network size $N$. We model social content filtering by implementing the limited visibility in the majority-vote model opinion dynamics, emulating how social media platforms decide which content to show users. Individuals are subject to content filtering in which each neighbor has a probability $V$ of being seen or filtered out during an interaction attempt. The social network experiences a social noise $q$, representing the inconformity probability. After considering the opinion of the visible neighbors, individuals have a probability $q$ of disagreeing with the majority opinion and a probability $1-q$ of agreeing with it.

We employ Monte Carlo simulations and estimate several parameters to analyze consensus in different configurations of $z$ and $V$. Our results show that increasing the social noise $q$ induces a second-order phase transition from consensus to dissensus at a critical value $q = q_{c}$. We show that stronger filters reduce the ability to form consensus in society and that increasing the number of connections each individual has increases the ability to maintain consensus. The critical noise $q_{c}(V,z)$ strongly depends on the visibility parameter $V$ and the growth parameter $z$ increases, in which we observe that high values of $z$ and $V$ yields high values of $q_{c}(V,z)$. Additionally, the finite-size scaling analysis also validates the unitary relation for critical exponents under the volumetric scaling of opinionization and noise susceptibility.

 \noindent

Our investigation reveals the potential conditions social media users may experience due to filtering algorithms. Users can be influenced to consume specific content more frequently, shaping their views and creating opportunities to manipulate public opinion, impacting society in various social, political, environmental and economic aspects. However, filtering content can also diminish false information, fake news, dangerous content, and discriminatory material from social media, enhancing the platform's sociability and security. We highlight that open filtering policies may improve the overall impact of the online social community while building a reliable source of information and fair conviviality. Despite the secret nature of proprietary social media filtering technologies, we believe its disclosure can enhance the stable and peaceful development of modern society while encompassing expressive financial profits.

Further research on the effects of filtering algorithms in social media may consider other network frameworks and the extensive use of social media connection data. Different distributions of the visibility parameter, the presence of various types of agents and different opinion dynamics may also be considered, contributing to a deeper understanding of the impacts of social media on public opinion.

\section*{CRediT authorship contribution statement}
\label{sec:Credit}
\textbf{Giuliano G. Porciúncula:} Conceptualization, Data curation, Formal analysis, Investigation, Methodology, Software, Validation, Visualization, Writing original draft.
\textbf{Marcone I. Sena-Junior:} Funding acquisition, Writing original draft.
\textbf{Luiz Felipe C. Pereira:} Conceptualization, Formal analysis, Investigation, Methodology, Visualization, Writing original draft. 
\textbf{André L. M. Vilela:} Conceptualization, Data curation, Formal analysis, Funding acquisition, Investigation, Methodology, Project administration, Resources, Software, Supervision, Validation, Visualization, Writing original draft.

\section*{Declaration of competing interest}
\label{sec:Competing}
The authors declare that they have no known competing financial interests or personal relationships that could have appeared to influence the work reported in this paper.

\section*{Acknowledgments}
\label{sec:Acknowledgments}
The authors acknowledge financial support from UPE, FACEPE (APQ-1129-1.05/24, APQ-0565-1.05/14, APQ-0707-1.05/14, APQ-1117-1.05/22), CAPES (0041/2022), CNPq (306068/2021-4, 167597/2017, 309961/2017, 436859/2018, 313462/2020, 200296/2023-0, and 371610/2023-0). The Boston University work was supported by National Science Foundation Grant PHY-1505000. LFCP also acknowledges the visiting professors program at Sapienza Università di Roma.


\end{document}